\documentclass[a4paper,fleqn,usenatbib]{mnras}

\usepackage[T1]{fontenc}
\usepackage{ae,aecompl}

\usepackage{graphicx}	
\usepackage{amsmath}	
\usepackage{amssymb}	

\usepackage{float}
\usepackage{ulem}  
\usepackage{color} 

\newcommand{\gr}{\mbox{\,gr}}
\newcommand{\cm}{\mbox{\,cm}}

\newcommand{\kpc}{\mbox{\,kpc}}

\newcommand{\pc}{\mbox{\,pc}}
\newcommand{\Myr}{\mbox{\,Myr}}
\newcommand{\Gyr}{\mbox{\,Gyr}}
\newcommand{\kms}{\mbox{\,km s}^{-1}}
\newcommand{\pcc}{\mbox{\,cm}^{-3}}
\newcommand{\Msun}{\mbox{M}_\odot}
\newcommand{\muG}{\, \mu \mbox{G}}
\newcommand{\dyn}{\mbox{\,dyn}}

\title[MHD simulations of ram pressure stripping]{MHD simulations of ram pressure stripping of a disk galaxy}

\author[M. Ramos-Mart\'{\i}nez, G. C. G\'omez and A. P\'erez-Villegas]{
Mariana Ramos-Mart\'{\i}nez$^{1}$\thanks{E-mail: m.ramos@irya.unam.mx (MRM)},
Gilberto C. G\'omez$^{1}$
and \'Angeles P\'erez-Villegas$^{2,3}$
\\
$^{1}$Instituto de Radioastronom\'{\i}a y Astrof\'{\i}sica, Universidad Nacional Aut\'onoma
de M\'exico, Apdo. Postal 3-72, Morelia Mich. 58089, M\'exico\\
$^{2}$Max-Planck-Institut f\"ur Extraterrestrische Physik, Gie{\ss}enbachstra{\ss}e, D-85741, Garching, Germany\\
$^{3}$Universidade de S\~ao Paulo, IAG, Rua do Mat\~ao 1226, Cidade Universit\'aria, S\~ao Paulo 05508-900, Brazil }

\date{Accepted XXX. Received YYY; in original form ZZZ}

\pubyear{2017}

\begin{document}
\label{firstpage}
\pagerange{\pageref{firstpage}--\pageref{lastpage}}
\maketitle

\begin{abstract}
The removal of the interstellar medium (ISM) of disk galaxies through ram
pressure stripping (RPS) has been extensively studied in numerous simulations.
Nevertheless, the role of magnetic fields (MF) on the gas dynamics in this
process has been hardly studied, although the MF influence on the large-scale
disk structure is well established. With this in mind, we present a 3D
magnetohydrodynamic (MHD) simulation of face-on RPS of a disk galaxy to study the
impact of the galactic MF in the gas stripping. The main
effect of including a galactic MF is a flared disk.
When the intracluster medium (ICM) wind hits this flared
disk, oblique shocks are produced at the interaction interface, where the ISM is
compressed, generating a gas inflow from large radii towards the central
regions of the galaxy. This inflow is observed for $\sim 150\Myr$ and may
supply the central parts of the galaxy with material for star formation
while the outskirts of the disk are being stripped of gas, thus the oblique
shocks can induce and enhance the star formation in the remaining disk. We also observed
that the MF alters the shape and structure of the swept gas, giving a smooth
appearance in the magnetized case and clumpier and filamentary-like morphology
in the hydro case. Finally, we estimated the truncation radius expected for
our models using the Gunn-Gott criterion and found that is in agreement with
the simulations.
\end{abstract}




\section{Introduction}

Lenticular galaxies (S0s) are objects that lie between the elliptical and spiral galaxies
in the Hubble sequence. The S0s share properties with both types of galaxies, that is, an old
stellar population like ellipticals and stellar disks like spirals. Lenticulars also have
prominent bulges \citep{sim86}, low gas content \citep{gal75} and some observations show that
the last star formation episode took place at the bulge (\citealt{pro11, sil06, sil12, bed12, eve12, eve14},
but see \citealt{kat15}).

The well studied environmental density-galactic morphology relationship in clusters of galaxies
\citep{dre80} states that late-type galaxies (spirals) are more frequently found in the outskirts
of clusters, while early-type {galaxies} (ellipticals and S0s) are more abundant in the central regions.
In the case of groups of galaxies, a similar trend has been observed \citep{pos84}. Additionally,
in cluster galaxies, the fraction of spirals {increases} with increasing redshift $z$, whilst the
S0s fraction decreases \citep{dre97,fas00}. On the other hand, when properties of spiral galaxies in
clusters and those in the field are compared (\citealt{bos06} and references therein), cluster
spirals are HI deficient and such deficiency increases towards the cluster {centre}. Also, cluster
galaxies show a lower star formation rate (SFR) associated with the lack of HI, and they are
redder than field galaxies, which indicates the former form stars passively \citep{byo78}.
Late-type galaxies also follow more radially elongated orbits than early-type, suggesting they
are free-falling into the cluster {\citep{gir86,dre86,vol01,biv04,agu17}}. Lastly, cluster
galaxies show an increase in radio-continuum emission, probably due to an enhancement in the magnetic
field (MF) intensity, possibly caused by compression \citep{sco93, ren97}.

These observations point to one or more mechanisms that act in the environment of clusters and
groups, stripping the galactic interstellar medium (ISM) from the disks or increasing its
consumption rate so that the star formation shuts down and a change in disk {colour} is produced.
Therefore, the idea that spirals are the progenitors of len\-ti\-cu\-lar galaxies has been proposed,
suggesting that the study of S0s may help us understand the impact of environment on the
evolution of disk galaxies.

In clusters, the main mechanisms proposed to explain the transformation of a spiral galaxy into an S0 are:

\begin{itemize}

  \item Ram pressure stripping (RPS, \citealt{gun72}): when a galaxy falls into the cluster {centre},
    the hot in\-tra\-clus\-ter medium (ICM) exerts an hydrodynamic pressure on the ISM of the galaxy and,
    if this pressure exceeds the gra\-vi\-ta\-tional force of the disk (Gunn-Gott criterion), then the
    ISM is stripped {off} the galaxy.

  \item Galaxy harassment \citep{moo96}: close and frequent encounters between galaxies occurring
    at high velocities, at a rate of one encounter per $1\Gyr$, may increase the SFR, rapidly
    exhausting the gas supply and eventually leading to a redder disk. These interactions will
    alter the galactic morphology by dynamically heating the disk.

  \item Starvation \citep{lar80}: the galaxy loses the envelope of hot gas that supplies the
    disk's gas reservoir, so the ISM is consumed and the star formation shuts down.

\end{itemize}

There are also other mechanisms that may act in groups of galaxies that can modify the galactic
 morphology, such as tidal interactions \citep{ick85}, and major \citep{too72, bor14}
and minor mergers \citep{agu01, tap14}.  Nevertheless, these processes are not
exclusive, that is, more than one might operate at the same time. Comparing {these} mechanisms,
\citet{bos06} conclude that RPS is the most appropriate to explain the differences observed in
between spirals of clusters and those in the field, since RPS removes the gas from the galaxies
producing a change in the SFR and colour. Also, RPS is efficient and inevitable near the cluster
centre and may alter indirectly the morphology of the disks (if a galaxy loses its gas, the stellar
disk is dynamically heated, leading to a thicker disk, {\citealt{far80,sel84,fvl98,bek02,elm02}}).

Multiwavelength observations have shown several cluster galaxies that are good candidates to be
experiencing RPS {\citep{koo04,chu09,yag10,ken14,bos14,bek14}}, since they show truncated
gaseous disks and in some cases gas tails, while the stellar disk remains unperturbed. \cite{cay90}
performed a survey of HI for spiral galaxies in the Virgo Cluster where they found that small
HI disks lie almost exclusively in the cluster centre in galaxies with high velocities with
respect to the cluster mean velocity, which make it possible they lost their gas through ram
pressure stripping. Moreover, they observed that galaxies affected by RPS have shown nuclear
activity. This could be since the gas pushed to the centre of the galaxy and the compression
exerted by the ICM enhances the star formation. Also, \cite{pog16} presented an atlas of galaxies
at low redshift that are being stripped of their ISM, with candidates found at all cluster centric
distances that showed an enhanced SFR compared to non-candidates of the same mass. This points
to the idea that RPS can induce and enhance the star formation.

A good example of a galaxy subject to RPS is NGC 4522 in the Virgo cluster. This is the most
studied case of a galaxy losing its ISM by this mechanism {\citep{vol00,vol04,vol06,vol08,
abr14,abr16,ste17}} and is possibly in the process of
transforming into an S0, since it shows a truncated disk in HI with a $3\kpc$ radius and a
$\sim 3\kpc$-length gas tail observed in HI \citep{keny04} and H$\alpha$ \citep{keny99}. Also,
in the Abell 3627 cluster, {the galaxy ESO 137-001 is stripped by the hot ICM
\citep{sun06,sun07,siv10,fum14,jac14,fos16}. ESO 137-001 presents} an $80\kpc$ long, double
X-ray gas tail \citep{sun06}, with some HII regions embedded within the tail \citep{sun10},
indicating that star formation can go on within the ISM stripped out of the galaxy. Later, in the same
cluster, another X-ray gas tail was detected (ESO 137-002), with a double H$\alpha$ tail \citep{zah13}.

The ICM-ISM interaction through the RPS has been studied extensively for years. A wide variety
of models have been developed with different me\-thods and techniques. The first models were
performed under the assumption of a constant ICM wind, u\-sing smoothed particle hydrodynamics
\citep[SPH;][]{aba99,sch01} and grid codes \citep{qui00,roe05,ro06a,ro06b}. These models were
in good agreement with the Gunn-Gott estimation for the disk
truncation radius. Other simulations were done varying the
inclination angle of the disk with respect to the wind direction \citep{vol06,ro06a,jac09}.
Yet other models added a variable ICM wind, so the RPS mechanism is not constant \citep[with
a sticky-particle code]{roe07,roe08,vol01}. Another extension to the RPS models included a
multiphase gas disk \citep{qui00,ton09,ton10}, where the low-density gas is stripped more easily
from the galaxy, but the mass loss of the ISM is not so different from homogeneous disk models.
Some other works included star formation, which showed an increase in the star formation in
central regions of the target galaxies \citep{sch01,vol01} and sometimes stars were formed in
the gas tails \citep{bek03,kron08,kap08,ste12,ton12}.

Despite the huge variety of RPS models, there are very few including MF. MF
have been observed in galaxies from polarized emission, mainly in radio frequencies, and Faraday
rotation. MF in spirals have an ordered component, i.e. with a constant and coherent direction,
and a random or turbulent component that has been amplified and tangled by turbulent gas flows
(\citealt{bec05a}, \citealt{bec13} and references therein). Combining information obtained with
different techniques, it is possible to develop a model for the 3D structure of MF in galactic
disks. In spirals, the average total field strength is $\sim 9\muG$ \citep{nik95} and the regular
field strength is $1-5 \muG$ \citep{bec13}, in radio-faint galaxies like M31 and M33 the
total field is $6\muG$
\citep{gie12,tab08}, in gas rich spiral galaxies the total field is $20-30\muG$ \citep{fle11,fri16},
for bright galaxies $\sim 17\muG$ \citep{fle10}, in blue compact dwarf galaxies $10-20\muG$
\citep{kle91} and the strongest total fields are found in starburst and barred galaxies with
$50-100\muG$ \citep{ade13,chy04,bec05b}. Since the degree of polarization on average is low in the
spiral arms, the random field is assumed to be stronger, up to five times the intensity of the
ordered field, whilst in the interarm region the degree of polarization is higher, hence the
ordered field should dominate. Additionally, it has been observed that the ordered MF shows a spiral
pattern that is offset from the spiral arms of gas and stars \citep{bec05a}.

\cite{rus14} presented simulations of RPS with a magnetized ICM and found that the MF can affect the
morphology of the stripped gas tail, since they observed narrower tails than in purely hydrodynamic
(HD) simulations. \cite{pfr10} also showed magnetohydrodynamics (MHD) simulations in which the galaxies
are moving in a magnetized ICM. The galaxies in their simulations swept the field lines where polarized
radiation is generated. This is used to map the orientation of the MF in clusters, e.g. Virgo cluster.
In these cases, the MF has been implemented only in the ICM and not in the disks. Some examples of
models with magnetized disks are \citet{vol06,vol07,soi06}, where they used the method of \citet{otm03}
where the MF is evolved via the induction equation using a grid code with the velocity field of the
particles, so that the MF is advected with the gas. These simulations of RPS have been carried out first
with a sticky particle code, and then a toroidal configuration of the MF is given to the galaxy. Even if
the effect of the MF over the gas dynamics has not been taken into account, this method has been useful
to explain the polarized radiation in radio that is observed in some galaxies that may be affected by
the RPS, as in the case of NGC 4522 \citep{vol06}.

Additionally, \cite{ton14} performed MHD simulations for the RPS including galactic MF, but the ICM was
not magnetized. They found that MF do not alter or dramatically change the stripping rate of the gas
disk compared to pure HD simulations. Nevertheless, the MF have an impact in the mixing of gas
throughout the tail, since inhibits the mixing of the gas tail with the ICM, the unmixed gas survives
at large distances from the disk. Besides, the RPS may help magnetize the ICM up to a few $\mu G$.

Here, we present MHD simulations of ram pressure stripping of a disk
galaxy under the wind-tunnel approximation, for a face-on geometry.
Additionally, we performed two purely HD runs to compare with the
magnetized case and analyze the impact that the galactic MF has in
the stripping of the disk. In \S \ref{model} we present the
initial set up for the simulations, in \S \ref{results} we
describe the resulting gas and MF distribution, and in \S
\ref{conclusions} we discuss our conclusions.

\section[]{Model} \label{model}

We set up a magnetized disk in rotational equilibrium in a fixed gravitational
potential. We used the MHD code RAMSES \citep{tey02}, which is an adaptive mesh
refinement code, so we can have more refinement of cells in the desired
regions, and allows us to add MF in the simulations. The models were performed
in 3D with 11 refinement levels, for a resolution equivalent to $(2048)^3$
cells, in a box of $120\kpc$ in each direction.

\begin{table}
\begin{center}
\begin{tabular}{|l|c|}
\hline
{\bf Bulge} & ${\mbox M_1} =  1.39\times10^{9}\Msun$ \\ 
 & \, ${\mbox b_1} = 0.85\kpc$ \\ \hline
{\bf Disk} & \, ${\mbox M_2} =  1.62\times10^{10}\Msun$ \\ 
 & ${\mbox a_2} =  3.0\kpc$ \\ 
 & ${\mbox b_2} =  1.0\kpc$ \\ \hline
{\bf Halo} & \, ${\mbox M_3} =  6.96\times10^{10}\Msun$ \\ 
 & ${\mbox a_3} =   16.0\kpc$ \\ \hline
\end{tabular}
\caption{Length scale and mass parameters of the 
gravitational potential, as adjusted to approximate M33's rotation
curve. ${\mbox M_1}$ and ${\mbox M_2}$ represent the total
mass of the bulge and disk, respectively, while ${\mbox M_3}$ is a mass factor
for the halo, where its total mass is obtained up to a cutoff radius.}
\label{tab:pot}
\end{center}
\end{table}

\begin{figure}
  \includegraphics[width=0.45\textwidth]{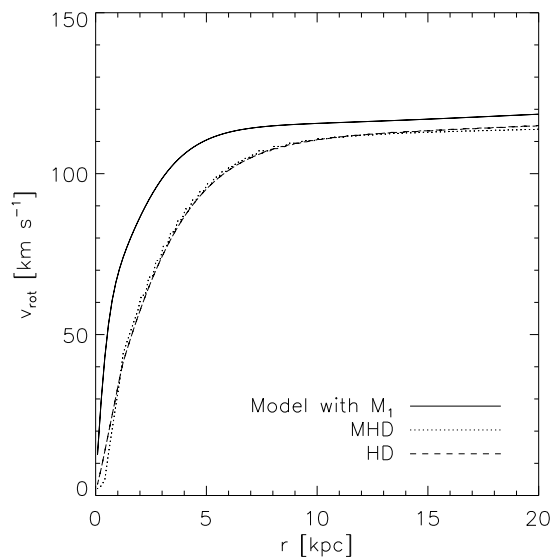}
  \caption{Rotation curve for the M33-like galaxy obtained from
  the gravitational potential with
  the values from Table \ref{tab:pot} {(solid line) and velocity profile
  for the MHD (dotted line) and the HD (dashed line) simulations without
  the bulge contribution ($M_1 = 0$)}.}
  \label{rotm33}
\end{figure}

\subsection{Initial Conditions} \label{sec:initmhd}

The gravitational potential used for our galaxy is based on the model of
\citet{potas91}, which is an analytic and simple potential that can reproduce
the rotation curve of the Milky Way and is composed by a spherical central bulge,
a Miyamoto-Nagai disk and a massive spherical halo. This potential model can be
easily modified to approximate the rotation curves of other galaxies. For this
work, we modeled an M33-like galaxy, which is a late type and low-luminosity
spiral galaxy. We modified the mass and scale parameters to the values shown in
Table \ref{tab:pot} to model the rotation curve of the M33 galaxy (see figure
\ref{rotm33}, {solid line}) as reported by \citet{cor03}. Nevertheless, for the
simulations presented in this work, we removed the galactic bulge component of
the potential ($M_1 = 0$) since it generated a large potential gradient in the
$z$-direction (perpendicular to the galactic disk) for small radii that generated
problems for our initial setup procedure (described below).
Regardless, this should have little impact on our conclusions, specially
since the M33 bulge's mass is small.
{The velocity profile used for the simulations, both with and
without magnetic fields, is also presented in figure \ref{rotm33}.}

For the initial conditions, we use a method similar to \citet{gom02}.
First, we define the radial density and velocity profile in the
galactic mid-plane,
assuming that the gas disk is in rotational equilibrium with the
gravitational force, the total pressure gradient, and the
magnetic tension,

\begin{equation}
  \frac{v_{\phi}^2(r,z)}{r} = \frac{\partial\Phi}{\partial r} + \frac{1}{\rho(r,z)}
       \left[ \frac{\partial P}{\partial r} + \frac{2P_B(r,z)}{r} \right] \, ,
  \label{balance}
\end{equation}

\noindent where the total pressure $P$ is the sum of the thermal
($P_{th} = c_s^2 \rho(r,z)$, with $c_s$ the sound speed) and the
magnetic ($P_B$) pressures.
The magnetic pressure has two components,
$P_B = P_{B,inner} + P_{B,outer}$,
with

\begin{eqnarray}
  P_{B,inner} &=& P_{B0} \left[1 - {\mbox {erf}} \left(\frac{R}{r_b} \right) \right] \quad\mbox{and} \label{Pbinn}\\
  P_{B,outer} &=& \frac{P_{B0}\,n}{(n+n_c)} \, ,
  \label{Pbout}
\end{eqnarray}

\noindent
where $R = \sqrt{r^2 + z^2}$, $r_b = b_1/3$ (see table \ref{tab:pot}),
$P_{B0} = 1.75 \times 10^{-13}\dyn\cm^{-2}$ and $n_c = 0.04 \pcc$.
With these expressions for the total pressure and eq. (\ref{balance}),
a given midplane density (or velocity) profile uniquely defines the velocity
(density) profile.
In the bulge, the rotation curve resembles a rigid body,
and so we define the rotation velocity linearly increasing with
radius,

\begin{equation}
  v_{\phi}(r,0) = v_{\phi}(b_1,0) \left[ \frac{r}{b_1} \right],
\end{equation}

\noindent where $v_{\phi}(b_1,0)$ is the circular velocity obtained from the
gravitational potential in $r=b_1$ and $z=0$.
Then, the density profile in the midplane is given by,

\begin{equation}
  \frac{\partial \rho}{\partial r} = \frac{ \rho \left( v_{\phi}^2
                     - \partial\Phi/\partial r \right)
                     + \left( P_{B0}/r_b \right) e^{-(R/r_b)^2} }
                     {\left[c_s^2 + P_{B0}\rho_c/(\rho + \rho_c)^2\right]}\, ,
  \label{drhodr}
\end{equation}

\noindent
which is integrated from $r=0$ to $b_1$.

For $r>b_1$ we do the converse: we define the density profile
as exponentially decreasing at the mid-plane,

\[
 \rho(r,0) = \rho_0 \, {\mbox{exp}[-(r-b_1)/h_r]}\, ,
\]

\noindent
where $h_r = 6\kpc$ and $\rho_0$ is the value found at $r=b_1$ from eq.
(\ref{drhodr}).

Once the mid-plane density is calculated, the distribution at $z\ne
0$ is found by assuming magnetohydrostatic equilibrium {and an isothermal
equation of state},

\begin{equation}
  \frac{\partial P}{\partial z} = - \rho \, \frac{\partial \Phi}{\partial z} \, ,
\end{equation}

\noindent
where again $P = P_{th} + P_B$.
By substituting the magnetic pressure components (eqs. \ref{Pbinn}
and \ref{Pbout}) and the equation of state it follows

\begin{equation}
  \frac{\partial \rho}{\partial z} = \frac{ -\rho \partial\Phi/\partial z
                     + \left(P_{B0}/r_b\right) \exp{[-(R/r_b)^2} ]}
                     {\left[c_s^2 + P_{B0}\rho_c/(\rho + \rho_c)^2\right]} \, ,
  \label{drhodz}
\end{equation}

\noindent
which is integrated along the $z$ coordinate to obtain the vertical
density profile at any radius $r$.

The rotation velocity above the midplane is given by \citep{gom02}:

\begin{equation}
  v_{\phi}^2 (r,z) = v_{\phi}^2(r,0) - v_A^2(r,0) + v_A^2(r,z) \, ,
  \label{vrotz}
\end{equation}

\noindent where $v_A$ is the Alfv\'en velocity ($v_A = \sqrt{2P_B / \rho}$).

\begin{figure}
  \includegraphics[width=0.5\textwidth]{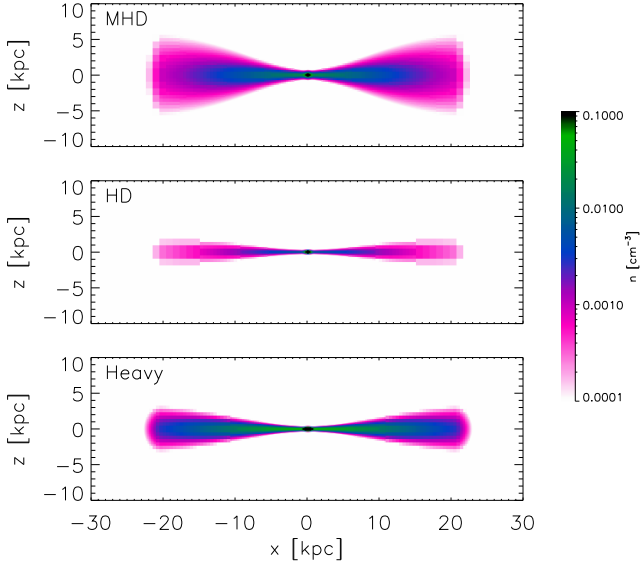}
  \caption{Density slices of the initial condition for the MHD ({\it top}), the HD ({\it center})
  {and the heavy ({\it bottom})}
  models. {Colours} show the gas number density at the $y=0$ plane.
  }
  \label{mapa}
\end{figure}

Figure \ref{mapa} shows a density map for the initial condition of the disk at $y=0$,
both for a magnetized and a purely HD ($P_{B0}=0$) cases. It can be seen that, for the
magnetized case, the disk is thicker in the outskirts than in the central region, that
is, the galactic disk flares in the presence of the MF. Additionally, the scale height
of the MHD disk is larger than in HD one. { When solving the equation of hydrostatic
equilibrium, the MF changes the compressibility of the gas,
thus increasing the surface density $\Sigma$ for the gravitational
potential and midplane density used, which results in a heavier disk compared to the HD
model. For this reason, we performed another HD simulation with
surface density similar to that in the magnetized disk model. We will refer to this
as the heavy disk model. To obtain the density distribution of
the heavy disk, we solve again the equations (\ref{drhodr}) and (\ref{drhodz}), increasing
the initial value of $\rho$ one order of magnitude over
original HD model, {thus increasing $\rho_{HD} (z=0)$ results in $\Sigma_{heavy} \sim
1.5 \Sigma_{MHD}$}. The initial density distribution
for the heavy model is also presented in figure \ref{mapa}. }

\subsection{Galactic magnetic field}

In the setup described above {for the MHD model},
the MF has two components (eqs. \ref{Pbinn} and
\ref{Pbout}). While the outer component ($r>b_1$) is purely toroidal, the inner one is
random. For the random inner component ($r< b_1$), we defined the vector potential
${\bf A}$ with,

\begin{eqnarray}
  A_x &=& A_0 \cos{\phi_r} \, \sin{\theta_r} \, f(z) \\
  A_y &=& A_0 \sin{\phi_r} \, \sin{\theta_r} \, f(z) \\
  A_z &=& A_0 \cos{\theta_r} \, f(z) \, ,
\end{eqnarray}

\noindent where the angles $\phi_r$ and $\theta_r$ where obtained randomly
and $A_0$ is drawn from a normal distribution with dispersion equal to
$\sqrt{8\pi P_{B0}}$.

\begin{figure}
  \includegraphics[width=0.5\textwidth]{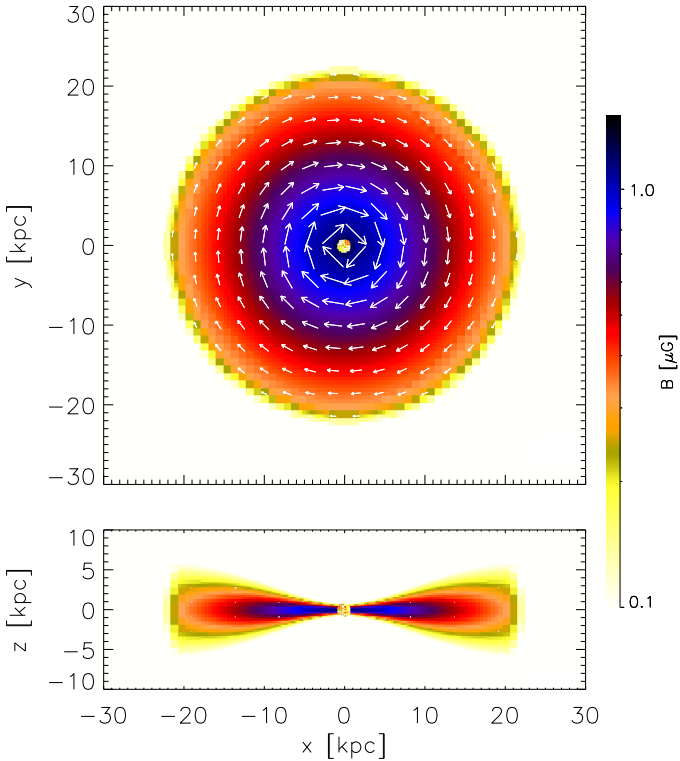}
  \caption{Slices of the MF intensity for the MHD model at the initial condition
           in a $z=0$ (top) and $y=0$ (bottom) cut. The MF strength is {colour}-coded
           in log-scale.}
  \label{Bmag}
\end{figure}

The function $f(z) = {\rm sech}^2(z/z_h)$, with $z_h = 150\pc$, modulates
the vector potential so its magnitude has the same scale height as the
density in the bulge. Once the components for the vector potential are
calculated, it is smoothed in order to avoid large fluctuations. Finally,
the MF is calculated ${\bf B}_{inner} = {\bf \nabla} \times {\bf A}$. 
For the rest of the disk ($r>b_1$), the MF in the setup follows
a toroidal configuration, with its strength given by the gas density (eq.
\ref{Pbout}). Figure \ref{Bmag} shows the initial intensity of the MF with
arrows overlaid representing the field lines for the MHD model.

\subsection{ICM wind} \label{sec:wind}

To simulate the ICM-ISM interaction, we worked under the wind-tunnel approximation, this
is, we place the galaxy at rest and the ICM flows towards the disk face-on. {The
ICM wind is unmagnetized and has the same parameters for all models: the wind starts at
$z=-10\kpc$ and moves in the $+z$ direction with density $n_{\rm ICM} = 10^{-5}\pcc$
and a velocity that increases linearly in time, from $300\kms$ to
$760\kms$ at the end of the simulation, at $500\Myr$.}
All the computational boundaries are outflowing, except at the
bottom where the wind flows inward.


\section{Results} \label{results}

\subsection{Model evolution} \label{evolution}

\begin{figure*}
  \includegraphics[width=0.99\textwidth]{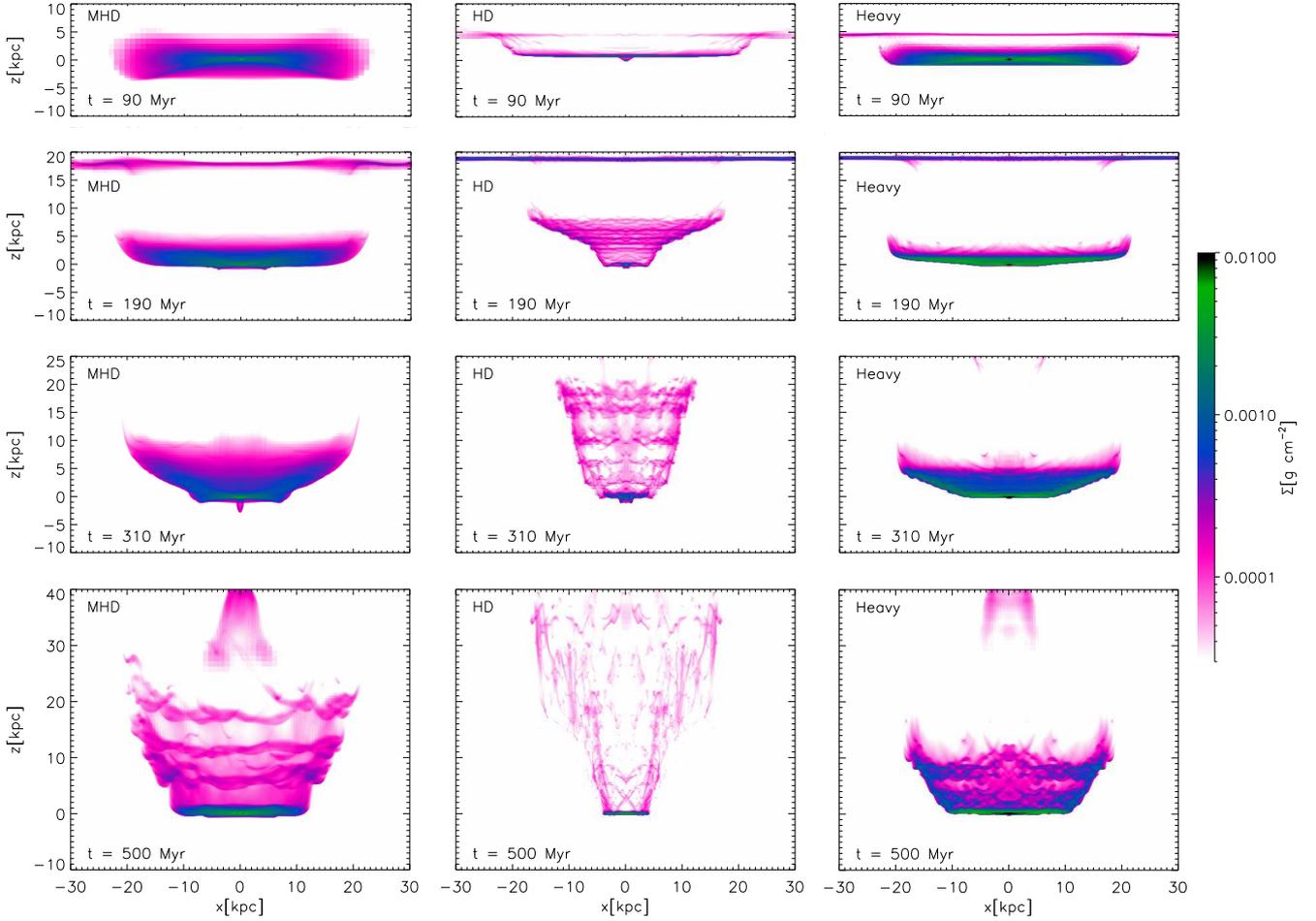}
  \caption{Projected density of the gas along the $y$-axis for the MHD ({\it left column}),
  {HD ({\it center})} and {heavy} ({\it right column}) runs. Each row corresponds to times $t = 90, 190, 310$ and
  $500 \Myr$. The ICM wind moves up from bottom of the computational grid.
  {Please note that the $z$-axis limits are different for
  each time shown.}
  }
  \label{fig:evol}
\end{figure*}

Figure \ref{fig:evol} shows the evolution of the models: {the magnetized
and both HD and heavy hydro models}, in maps of projected density.
The first row corresponds to a time $t = 90 \Myr$. It can be seen
that the wind is starting to interact with the disks. In the MHD
run (left column), oblique shocks appear on the side of the disk
that is facing the ICM wind (at $z<0$), because our disk flares
in the presence of the galactic MF, giving  it a
``bow tie'' shape. The oblique shocks lead gas of the external
parts of the disk towards the galactic {centre} and continues for
another $\sim150\Myr$ more until the ICM wind finally surpasses
the gravitational force of the disk and starts to sweep the ISM
(see \S \ref{sec:shock}). Since the HD disk ({central} column)
does not flare as much as the MHD one, the ISM-ICM interaction
is different. {At $90\Myr$, in the HD model} the wind perturbs the gaseous
disk and displaces it from the  $z=0$ midplane. The background
gravitational force  pulls back the gas to its original position,
mainly in the inner region of the disk, whilst the outer parts of
the disk are still being swept by the wind. {In the heavy model,
most of the gas disk initially at $z<0$ is compressed and moved by the wind 
to a height $z\sim 0$, changing the disk symmetry but to a smaller
extent than in the HD case.}

At $t= 190\Myr$ (second row), for the MHD run, most of the gas that lied below
the galactic midplane was swept by the wind which in turn starts to erode the
disk at large radii, where the ram pressure exceeds the gravitational force of
the galaxy (see \S \ref{sec:gg}). The inner region of the disk ($r<5\kpc$) is
slightly perturbed, with small variations in the $z$ direction, since the
gravitational force tries to keep the disk in the equilibrium position. As a
result, the gas moves up and down. These fluctuations in $z$ at small radii
occur at earlier times in the HD model but it is basically the same behavior:
gas at large $r$ is swept by the wind whilst the gas located near the galactic
centre remains bound to the disk. The displaced gas reaches a height of
$\sim 5\kpc$ and $\sim 10\kpc$ above the galactic midplane, for the MHD and the
HD case, respectively.  The HD model shows a larger erosion than the MHD, with a
gaseous disk of radius $r < 5\kpc$ for the HD run, and $r \sim 20\kpc$
for the MHD one.  The gas of the heavy disk has moved just a few $\kpc$
from the midplane, showing almost the same radial extension {as the MHD},
although the heavy disk is denser near the galactic midplane.

After $310\Myr$ of evolution, the wind continues flowing and
accelerating towards the disk and reaches $v_{\rm ICM}\sim 550\kms$
at $z=0$. For the magnetized disk (third row, left), the gas at
$r>10\kpc$ is ripped off of the galaxy, where the ram pressure is
stronger than the galactic gravitational force. The swept gas has
increased its height, reaching
$z \sim 10\kpc$. There are still some vertical motions in the midplane
for $r<10\kpc$ because the gas in this position is adjusting to the
balance between the pressure from the wind and the gravitational
force in the disk. This process is also present in the HD run,
but the oscillating gas is contained in a smaller
radius ($r<7\kpc$). Additionally, for the HD model, the gas
that has been removed from the disk has reached a height of
$\sim 20\kpc$ above the disk midplane. Compared to the
magnetized case, the stripped gas has a more diffuse appearance in
the HD run, that is, the gas mixes easier with the surrounding,
and it is less extended in the radial direction
than the MHD case. The HD disk is, at this time, truncated to a
radius of $\sim 6\kpc$.
The heavy disk at $t = 310\Myr$ shows a structure similar to the MHD
one: the heavy disk has a radial extension of $r\sim 10\kpc$ in the midplane,
while in the vertical direction the denser gas reaches a height of $z\sim 5\kpc$,
{but the less dense gas is farther away the galactic midplane} in the MHD model
($z\sim 10\kpc$).

At $t=500\Myr$ the wind has a velocity of $\sim 760\kms$ at $z=0$ for all cases.
In the MHD model, the stripped gas reaches a height of $\sim 20\kpc$ above the
galactic midplane. The disk has been truncated to a radius of $\sim 10\kpc$,
which is approximately half of its original size. The  dimensions of the
displaced gas for the MHD model resembles the one from the HD at $t = 310\Myr$,
showing a similar longitude over the midplane, which suggests that the evolution
of the MHD simulation is delayed {with respect to} the HD run, although
differences remain in the morphology: in the MHD case, the swept gas has a
smooth appearance and is denser at higher $z$ than in the HD case, which indicates
that the MF prevents the gas from mixing with the surroundings, similarly as
seen in \cite{ton14}. On the other hand, the HD model with $500\Myr$ of
evolution shows a more filamentary and clumpier morphology in the stripped gas,
contrary to the smooth appearance that the magnetized gas presents. The HD gas
is extended over $\sim 40\kpc$ above $z = 0$, and the remaining disk has radius
of $\sim 4\kpc$, which indicates that this disk has reached a state of
equilibrium with the ram pressure, since the gas was rapidly eroded in the first
$\sim 200\Myr$ of evolution and the remaining gaseous disk (in $z = 0$) has the
same radial extension until the end of the simulation. The heavy disk
has a size similar to the MHD ($r \gtrsim 10\kpc$) in the midplane ($z=0$),
suggesting that the stripping rate for both disks is approximately the same,
whilst the displaced gas for the heavy model has a lower $z$-height.
Nevertheless, when the heavy and the HD simulations are compared, the displaced
gas of the heavy model resembles the HD case, in that both have a clumpy and
filamentary-like structure, with the difference that, in the heavy model, the
swept gas is denser {because of the initial condition of the gas disk,
that is $\rho_{heavy} > \rho_{HD}$ ($\Sigma_{heavy} > \Sigma_{HD}$) as mentioned
in \S \ref{sec:initmhd}, which also results in a slower erosion of the disk}.

Comparing the evolution of the three models, the MHD and the heavy
model are left with a similar remnant disk, with radius $r\sim 10-12\kpc$ which
is larger than the HD model ($r \sim 4\kpc$) for the same time of evolution.
Our results suggest that the stripping rate depends on the MF only
through the surface density $\Sigma$ of the disk: a heavier disk (high $\Sigma$)
is more difficult to erode since the ICM has more material to sweep, even when
the gas is {farther} away from the gravitational potential well, and thus less
bound to the galaxy. This is similar to the results presented by \cite{ton14},
where the MHD and the HD disks with the same initial mass do not show a
significant difference in the stripping rate. Although the heavy model agrees
quite well with the MHD in the rate at which the gaseous disk is removed and the
truncation radius (see \S \ref{sec:gg}), the problem with the heavy disk is that
it gives a higher and unrealistic volumetric density $\rho$ in the galactic
midplane, because in the absence of the MF, using the same potential to solve
the hydrostatic requires a high value of $\rho$ to obtain the same $\Sigma$ of
the magnetized case. Nevertheless this heavy model is useful to investigate the
dependence of the stripping with the disk surface density.

It is observed in our models that the MF has an impact in the morphology and
shape of the swept gas. In the magnetized case, the swept gas shows a smooth
structure with denser gas surviving at higher $z$, {similar to the results of}
\cite{ton14}; while in the two non-magnetized models, the gas located above the
midplane has a clumpy and filamentary shape. {The morphology of the swept gas in
the HD and heavy models is due to the equation of state of the gas. In the case where an
isothermal equation of state is implemented, like in the setup we presented, the gas is
more compressible compared to an adiabatic or magnetized gas (with an adiabatic index
$\gamma > 1$). In our isothermal models, when the wind hits the galaxy, the gas disk is
compressed so that clump-like regions form, leading to the development of eddies due to
instabilities in the gas and when the eddies are pushed upwards by the wind they
generate a tail. This behavior of the gas is similar to the flow of the cigarette smoke,
giving the filamentary and clumpy shape to the swept gas in our HD simulation.}

It is noticeable that, in some aspects, the swept gas in our MHD model resembles
the HI distribution of the spiral galaxy NGC 4522, a galaxy considered a classic
example of RPS (see the figure 2 from \citealt{keny04}): the HI distribution is
asymmetric with respect to the stellar disk, is cap shaped, the gas contours are
compressed in the upstream side, and it is concave or curved to the downstream side.
On the other hand, {the stripped gas is not as far from} the NGC 4522 disk
as the gas distribution in our MHD model at $t = 500\Myr$.

\subsection{Gunn-Gott criterion} \label{sec:gg}

\begin{figure}
  \includegraphics[width=0.45\textwidth]{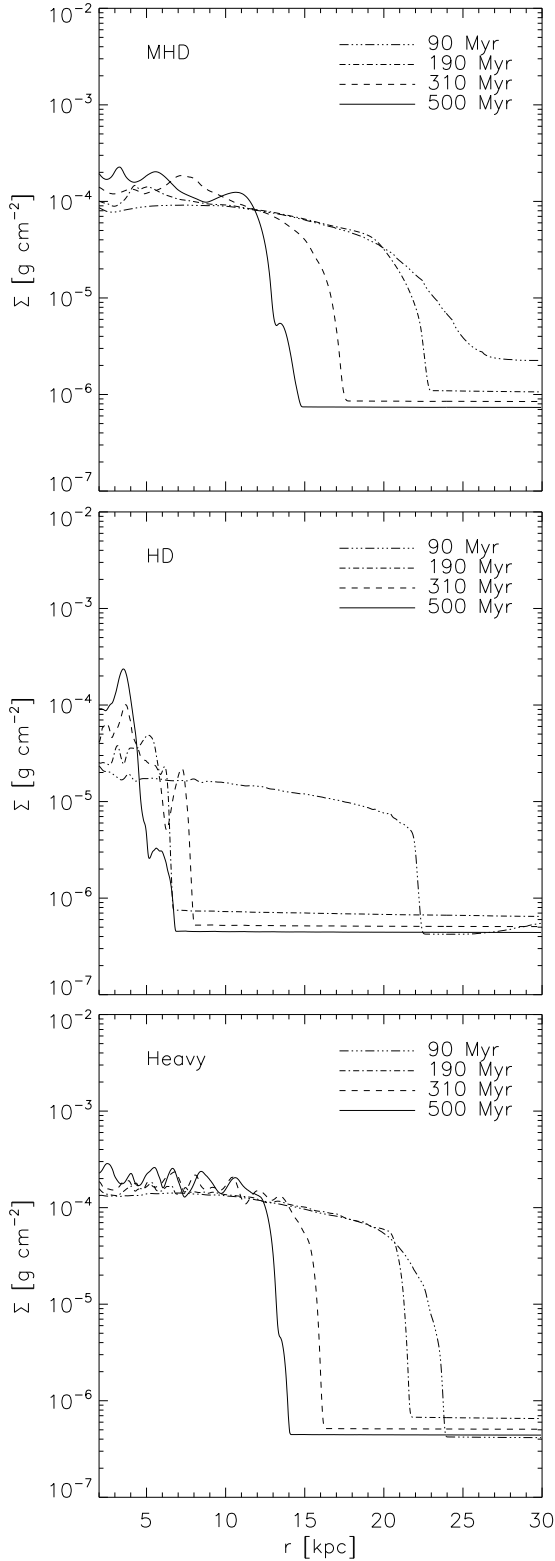}
  \caption{Evolution of surface density for the MHD ({\it top}),
  the HD ({\it middle}) {and the heavy ({\it bottom})} models as a function
  of galactocentric radius.
  The surface density is calculated up to $z = \pm 5\kpc$ from the midplane for the
  MHD disk and $z = \pm 3\kpc$ for the HD and heavy disks. The MHD {and heavy}
  disks ({\it top} and {\it bottom}) are eroded more slowly than the HD ({\it middle}).
  }
  \label{fig:supdens}
\end{figure}

The Gunn-Gott criterion (GG, \citealt{gun72}) estimates the
radius at which a disk galaxy, experiencing the ram pressure
face-on, will be truncated. This is determined by equalizing the
ram pressure $P_{\rm ram} = \rho_{\rm ICM} v^2_{\rm ICM}$ exerted by
the wind and the gravitational restoring force in the disk, which
is the product of the gravitational force of the galaxy and the
surface density of the gas disk $F(r)\Sigma(r)$, that is, the
truncation radius is defined by the position where
$P_{\rm ram} = F(r)\Sigma(r)$.

\subsubsection{Disk surface density}

In order to verify that our simulations satisfy the GG criterion, we estimate the
truncation of our disks measuring the surface density in the $z$ direction. We
did {these calculations} over time also to study the differences in the stripping
rate for the three models. Figure \ref{fig:supdens} shows the
evolution of the disk surface
density ($\Sigma$) over time obtained for $|z| \leq 5\kpc$, 
for the MHD model, and $|z| \leq 3\kpc$ for the HD and heavy disks. The differences
in the $\Sigma$ integration range in the $z$-direction are due to the different
thickness of the MHD with respect to the HD and the heavy ones.
We define the truncation radius as the one where $\Sigma$ decays
abruptly. At $t = 90\Myr$, the disks are barely perturbed, as can be
seen by comparing with figure \ref{fig:evol}, and their surface density distribution is
similar to the initial condition: the surface density decreases slowly in $r$,
and decays rapidly at $r \sim 21-22\kpc$, except for the MHD disk where
the decay is less abrupt. Given that in the MHD
case $\Sigma$ decreases approximately two orders of magnitude
($10^{-4}-10^{-6}\gr \cm^{-2}$) in $r > 20\kpc$, to obtain the truncation radius of the disk we
took the midpoint for this range of densities, in the log-scale, and then we found
the value of $r$ where we have this density ($\Sigma = 10^{-5}\gr \cm^{-2}$),
giving $r\sim 23\kpc$. As mentioned before (see figure \ref{mapa}), the MHD disk has a
higher surface density than the HD disk by approximately one order of magnitude
due to the extra support that the MF provides. By construction, {the heavy disk
has a value of $\Sigma_{heavy} \sim 1.5\Sigma_{MHD}$} (see \S \ref{sec:initmhd}),
but the decay of $\Sigma_{heavy}$ is more abrupt than the MHD case and lies between
the range of $r = 21-22\kpc$, similarly to the HD model.
At $t = 190\Myr$, the surface density in the inner MHD disk is still similar to
the {previous snapshot, but $\Sigma$ decreases more rapidly with radius
than at $90\Myr$, resulting in a disk with $r\sim 19\kpc$}, which is a
clear sign that the ICM {has started to erode} the gas of the disk. At this
same time, for the HD run, the disk has been eroded more efficiently than the
magnetized one, with $\Sigma_{HD}$ showing an abrupt drop at $r \sim
6\kpc$. The heavy disk shows a truncation radius of $r \sim 20\kpc$
and is also evolving similarly to the MHD run.

When the simulations have reached a time of $310\Myr$,
is clear to see that the surface density profile has changed for the MHD and heavy disks,
due to the accelerating wind that swepts the gas of these galaxies.
This is observed in the fall of the density and the smaller radial extension of
the disks, which have been reduced considerably to $r \sim 15\kpc$
for the MHD and heavy models.
The HD model evolves faster in time than the other models, as expected,
since most of its disk was swept at earlier times ($t=190\Myr$), presenting a
gaseous disk with $r\sim 7\kpc$.

The ICM wind keeps eroding the gaseous disks of the three models until
the end of the simulation ($t = 500\Myr$), leaving a remnant disk with $r \sim 12
\kpc$ and $r = 10-12\kpc$ for the MHD and heavy models, respectively.
The HD simulation was run longer, but
the disk length reaches an approximate steady truncation radius of $\sim 4\kpc$ at
$t=500\Myr$, showing that the erosion of this model was faster and more
efficient than in the MHD and heavy disks, that loose their gas at a slower rate,
as mentioned in \S \ref{evolution}, and whose disks are truncated at a larger
radius.

\begin{figure}
  \includegraphics[width=0.5\textwidth]{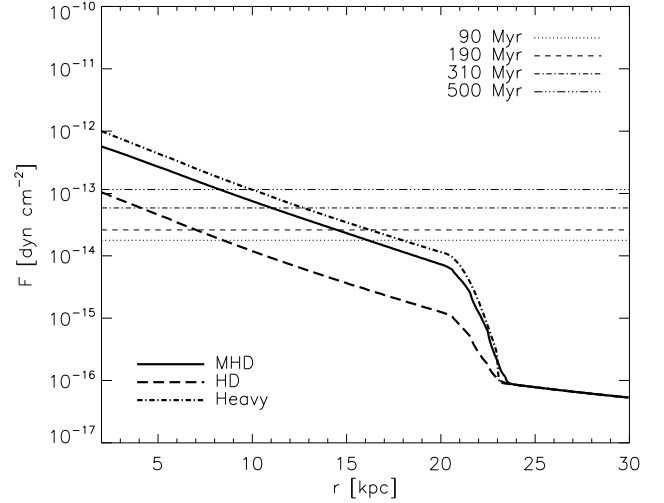}
  \caption{{The Gunn-Gott criterion \citep{gun72} for disk gas
  stripping applied to the simulation disks.}
  The thick lines represent the restoring force of the MHD
  (solid line), the HD (dashed line) {and the heavy (dot-dashed line)}
  disk, while the horizontal lines
  correspond to the ram pressure measured at different times. For a given set of
  wind parameters, the radius where the $P_{\rm ram}$ intersects the
  {restoring force defines}
  the truncation of the gas disk.
  }
  \label{fig:fgrav}
\end{figure}

{It is worth mentioning that the increase in $\Sigma$ is related with
the following numerical factors. First, is the difficulty of modeling a
Cylindrical system in a Cartesian grid. The gas fluxes across grid boundaries
in this mismatch lead to errors when the curvature of the circular orbits is
large, generating spurious radial flux and a lack of proper rotational support.
Second, the rapidly changing gravitational potential in the central regions of
the galaxy. The HD disk has a scale height of $\sim 200\pc$ or even smaller at
$r=0$, and with the best spatial resolution achieved just a few grid points
are calculating the hydrostatic. We tested how much our models deviate from 
equilibrium by performing simulations of isolated MHD and HD disk and found
that the ill-resolved hydrostatics and rotation generates a collapse of material
in the centre of the galaxy, which yields to an increase in the surface density.
In the HD isolated disk, the surface density increases from 1 to 2 orders of
magnitude in $r<2\kpc$ from $t=0$ to $t=500\Myr$. There is also an infall of
material in the isolated MHD model, but since this disk is more extended in the
z-direction, the increase in the surface density is less than one order of
magnitude for the same radii and time of evolution compared with the HD case,
this is because the grid effects are smaller in the MHD model. When the wind is
on, the increase of the surface density is lower than in the isolated cases,
since the interaction with the wind diminish this effect. With this in mind, 
the surface density is not an adequate measure of the inflow of gas derived
from the oblique shocks in our models (see \S \ref{sec:shock}).}

\subsubsection{Disk truncation}

Figure \ref{fig:fgrav} shows the gravitational force per unit area for {the
MHD (solid thick line), the HD (dashed thick line) and the heavy (dot-dashed thick
line) disks}, approximated as follows: using the gravitational potential of
{the background} axisymmetric model we obtained the maximum force in the
$z$ direction as function of $r$ and multiplied by the surface density $\Sigma(r)$:

\begin{equation}
  F(r)
  = F_{z \rm max}(r) \, \Sigma(r) = -\frac{\partial \Phi(r,z_{\rm max})}{\partial z}
                      \, \Sigma(r) \, ,
  \label{ggradius}
\end{equation}

\noindent where 
$z_{\rm max}$ is the point where the gravitational force is maximal. {Notice that
the gravitational potential is the same for all models but the differences in the
restoring force are due to the initial surface density in the disks (see \S \ref{sec:initmhd}).}

This gravitational restoring force is compared with the ram pressure $P_{\rm ram}
= \rho_{\rm ICM} v^2_{\rm ICM}$ exerted by the wind (represented with the horizontal
lines in the figure). The gravitational force of the disk decreases with
increasing radius so, for a given set of parameters for the wind, we expect that
the disks are truncated at the radius where both forces are equal, that is, where
the $P_{\rm ram}$ and the force lines cross each other. For the wind parameters,
we have $n=10^{-5}\pcc$ and the velocity is taken from the simulation. Since it
increases in time we chose the value of $v_{\rm ICM}$ at $z=0$, when it has
reached the disk midplane. The lines for the ram pressure are labeled
according to the time at which the wind velocity was calculated.

\begin{figure}
  \includegraphics[width=0.5\textwidth]{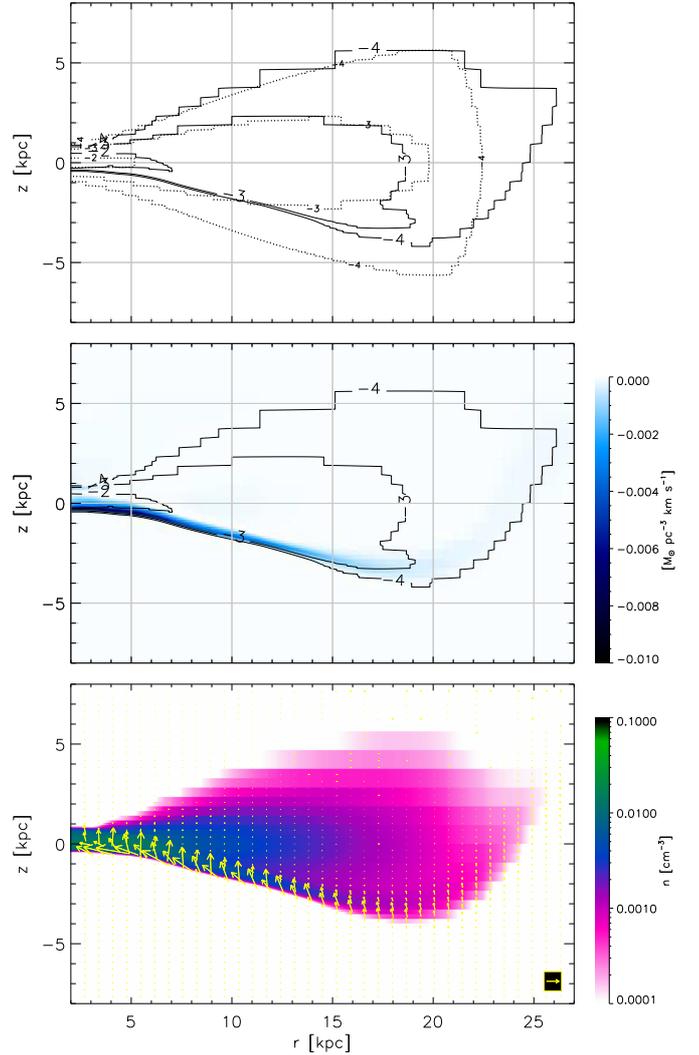}
  \caption{{\it Top}: Slice of density for the MHD disk in a $y=0$ cut at time
  {$t=0 \Myr$ (dotted line) and} $t=90\Myr$ {(solid line)}.
  The contour levels are $n=10^{-4},10^{-3}$ and $10^{-2} \pcc$
  and are labeled from outside-in.
  {\it Middle}: Radially {inward} flux of mass due to the oblique shocks at
  the ICM-ISM interaction zone for the same timestep. {Colours}
  in the flux map show radially inward mass flux, {while the
  contours are the same as the ones for $t=90\Myr$ in the top panel, for
  comparison.}
  {\it Bottom}: Slice of density for $t=90\Myr$ in a $y=0$ cut,
  where the colours show the gas number density and flux arrows are
  overlaid. {The scale in the bottom right box corresponds to a mass
  flux} of $3 \times 10^{-3} \Msun \pc^{-3} \kms$.
  The oblique shocks are produced (due to the ``bow tie'' shape of the disk)
  {at the disk-wind interface}
  and move the gas {from the outer galaxy}
  towards the galactic {centre}.}
  \label{fig:flux}
\end{figure}

Following the GG criterion, the truncation radius expected for the MHD,
HD, and heavy disks is $r \sim 16\kpc$, $\sim 8.5\kpc$ and $\sim 18\kpc$,
respectively, with the wind velocity measured at $t=90\Myr$, which is smaller
compared to the cut in the radial direction of our disks measured in the
simulation ($r_{MHD} \sim 23\kpc$ and $r_{HD,heavy} \sim 21-22\kpc$).
At $t = 190 \Myr$, the GG truncation radii 
are also smaller for the MHD and heavy disks compared to the ones
calculated from the simulation: $14.5\kpc$ ($19\kpc$ in the simulation) for
the MHD and $16.5\kpc$ ($20\kpc$ in the simulation) for the heavy model.  For
the HD disk, the truncation radius measured from the simulation is in better
agreement with the one predicted by GG, $r \sim 6-7\kpc$, and could be due to
the fact that this model loses its gas faster than those with a higher initial
$\Sigma$.

At later times, from $t=310$ to $500\Myr$, the truncation radius
from GG is more similar to the observed in the simulations,
showing slight differences of $1-2\kpc$ in the two non-magnetized models.
By the end of the simulation, $t=500\Myr$, the radius of the MHD disk should
be $r \sim 8-9\kpc$ according to GG, while the
size measured is $r \sim 12\kpc$. For the HD model, GG predicts
$r\sim 2\kpc$ while we measure $r \sim 4\kpc$ in the simulation.
Finally, in the heavy disk we have $r \sim 10\kpc$ and $r \sim 10-12\kpc$ with
GG criterion and measured in the simulation, respectively.
{The three models are a reasonable fit to the GG criterion, although
the HD and heavy ones are marginally better. This} could
be due to the assumptions of GG: a zero-width disk (the HD model disk has a
scale height of $\sim 200\pc$) and no consideration of the effect of the
MF in the gas dynamics. Still, even if the values for the truncation
radius do not coincide exactly with the calculations from the simulations, the
GG criterion yields a good approximation of how much a gaseous disk may be
stripped due to ram pressure.

\begin{figure}
  \includegraphics[width=0.5\textwidth]{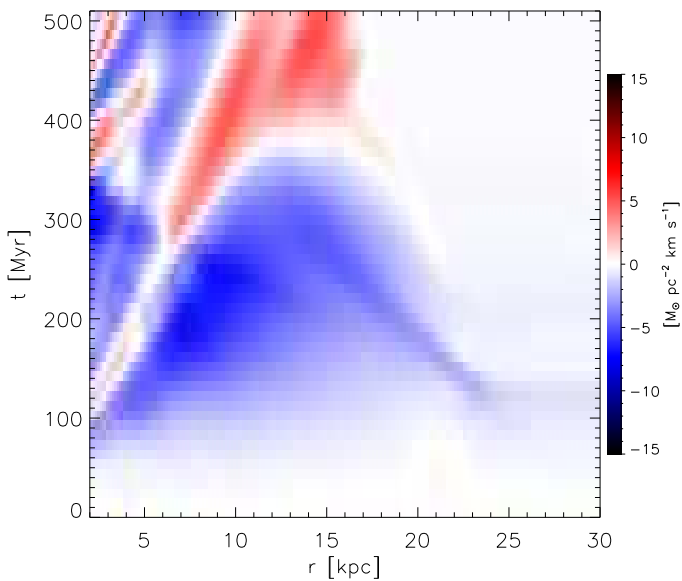}
  \caption{Flux of disk mass integrated in $z$ as a function of time.
  The radial flow azimuthally averaged is $z$-integrated within the range
  $|z| \leq 10\kpc$. The {colour}-bar shows the inward motions in
  blue and the outwards as red of the mass flux. The oblique shocks appear
  in $r = 5-10\kpc$ at $t \gtrsim 100\Myr$ and increase outwards,
  {driving} gas to smaller radii. The shocks {reach} their
  maximum strength at $t \approx 250\Myr$ and after that they start to
  vanish from the outskirts of the disk ($r>10\kpc$) when the ram pressure
  increases and generates an outward flow of gas instead at $t > 300\Myr$.
  }
  \label{fig:flux_in_time}
\end{figure}

\subsection{Oblique shocks}\label{sec:shock}

The MHD model has a flared disk (see \S \ref{sec:initmhd}) since the MF
yields a less-compressive gas layer. Therefore, when the ICM wind reaches
the galaxy, an oblique shock is generated in the wind-disk interface. The
shocks change the initial distribution of gas in the disk, as can be seen
in the density contours in figure \ref{fig:flux} (upper panel), and are
present in most of the wind-facing side of the disk. The figure compares
the initial distribution of the gas density of the disk (dotted line) and
at $t=90\Myr$ (solid line). The most diffuse gas with $n=10^{-4}\pcc$
(see the $-4$ contour) is pushed up and compressed so that the density is
more extended in the $+z$ direction than in the $-z$ direction.
Conversely, gas with $10^{-3}\pcc$ ($-3$ contour) is more extended in the
$-z$ direction compared to the initial distribution because of the
accumulation of the material due to the compression. {This
compression advances to smaller radii, so that there is more 
dense gas in the central regions of the galaxy, leading to an expansion
of the inner part of the disk with $n=10^{-2}\pcc$ ($-2$ contour) in the
radial direction and below the midplane in $5\kpc < r < 7\kpc$.}

These oblique shocks lead to a radial inflow of gas toward
the inner regions of the galaxy. The middle panel of figure
\ref{fig:flux} shows this (azimuthally averaged) mass flux at $t =
90\Myr$,
with the gas density contours at $t=90\Myr$ from the upper panel
are overlaid. Notice that the inward flux matches with the
shocked gas, that is the shocks funnel the gas from larger to
smaller radii.

The oblique shocks and the inward flux of mass they produce
are present at all radii, being stronger in the inner region of the
disk ($2\kpc< r< 10\kpc$) than in the external region. {The
inflow of gas can be also observed through flux arrows, which are
represented in the bottom panel of figure \ref{fig:flux}, where
these are overlaid on a density slice as in figure \ref{mapa}, in
the $y=0$ plane also for $t = 90\Myr$. The flux arrows show the
motion of gas to the centre of the galaxy produced by the oblique
shocks, as we mentioned earlier. The shocked gas at the interface
of the disk and the wind ($z<0$) is pushed up and redirected to
smaller radii.  These shocks and the inflow of gas from the
outskirts ($r>10\kpc$) may supply the central regions of the disk
to ignite star formation or nuclear activity, until the ram
pressure increases and starts to sweep the gas from the galaxy.}

Figure \ref{fig:flux_in_time} shows the evolution in time of the $z$-integrated
radial gas flux. The gas flux is integrated over a height of $|z| \leq 10\kpc$.
{Blue colour represents the radial inflow and
red colour is the outflow. As it was previously mentioned, the gas is compressed
and funneled to the inner regions of the galaxy. Both the shocks and the flow
appear at $t = 90-100\Myr$ and have a radial extension of $r = 6-8\kpc$, where the
flux is maximum at this time. As the
wind pushes a larger portion of the disk, the shocks and the inflow they produce
increase in radius as time increases. For example, in $15\kpc<r<20\kpc$ the
inflow is active from $t\sim 150\Myr$ to $t\sim 300-350\Myr$, which means that the
oblique shocks can funnel the gas from the outskirts of the disk to
smaller radii before the wind sweeps it out.}

The strongest inflow is generated in the time interval of
$\sim 100-250\Myr$ after the shocks appear, that is, in the
$t \sim 200-350\Myr$ mark in figure \ref{fig:flux_in_time}. After $t=300\Myr$,
the inflow from the outskirts, that is, the gas originally located in
$r>10\kpc$, becomes weaker until it starts to vanish at
$t\sim 400\Myr$. This happens when the wind accelerates and surpasses the galactic
gravitational potential, generating an outward flow instead and finally removing
the gas of the disk. The swept gas is represented by the red area near the end of
the simulation, $t>400\Myr$ and in the radial range of $r=10\kpc$ to $r=17\kpc$.
Even though at times $t > \sim 350\Myr$ for radii $r<10\kpc$,
there is still an inflow of gas to the centre of the disk, the motion
in this region is more random or disordered due to the interaction with
the high-speed wind ($v_{\rm ICM} > 700\kms$), which is observed in the
blue and red bands. Additionally, the initial flaring of the disk has
almost vanished since the wind has compressed and displaced the gas
below the midplane.

It can be seen from figure \ref{mapa} that the heavy disk is also flared but
to a lesser extent than the MHD one. {We also analyze the oblique shocks
in the heavy disk model.} We observed an increase in the density of the disk due
to a compression of the gas in the wind-disk interaction zone, {as in the
MHD model}. Nevertheless, the layer of the shocked gas is less prominent, with a
thickness of {$\sim 200\pc$ which is $\leq 0.4$ of the shocked region in
the MHD case}, where the thickness of the compressed gas ranges from
$\sim 500\pc-1\kpc$ in some regions of the disk. The intensity of the azimuthally
averaged flux {(as a function of $r$ and $z$) for the heavy disk has
nearly the same maximum value reached in the MHD case at the time shown in figure
\ref{fig:flux} ($t=90\Myr$),} but the strongest flux in the heavy model is
located near the centre of the disk, where our setup is not very reliable due to
the grid. Studying the flux arrows for the heavy disk we also observed that the
vertical motion of the gas in the $+z$ direction dominates over the radial one,
that is the wind mostly moves the gas upwards before funneling it to the centre
of the galaxy. {Addiotionally, the compressed layer of gas is closer to the
galactic midplane, so the shock is less oblique}. The inflow of gas as a function
time for the heavy model is on average a factor of $0.5$ lower than in the MHD
{since in the latter there is gas located at higher z and therefore, when
the flux is integrated in the z direction, there is more gas moving towards the
centre of the disk and the total flux as a function of time and r is higher. The}
strongest inflow in the heavy model is present between $t \sim 150-300\Myr$,
lasting $\sim 150\Myr$. At $t>300\Myr$ the motion of gas in the heavy model is
more disordered in the inner disk ($r<10\kpc$) until it is finally removed by the wind.

The inflowing gas driven by the oblique shocks raise the possibility of a strong
star formation episode in the central part of the galactic disk while the
outskirts of the galaxy are being stripped of gas. Observations suggest that S0
galaxies had their last star formation burst in their bulge (\citealt{pro11,
sil06, sil12, bed12, eve12, eve14}, but see \citealt{kat15}), so this mechanism
can provide the central regions with the gas necessary for that burst.
Additionally, galaxies undergoing RPS have shown unusual nuclear activity,
possibly because the gas is being pushed to the centre and also an enhanced
star formation in the region where the gas is compressed by the ICM, that is,
the star formation is induced and enhanced by the ram pressure \citep{cay90,
pog16}. \cite{pog17}, found a very high incidence of AGN (Seyfert 2) among
jellyfish galaxies from MUSE data and they conclude that ram pressure triggers
the AGN activity.

There are several points that need to be kept in mind when comparing our
simulations with the above quoted results. First, in this work we present only
a generic model for a flared disk galaxy. More studies must be performed in
order to verify the presence of these oblique shocks in galaxies. In our model,
the flare is created by a magnetic field, but this is not the only mechanism
to create such a disk (for example, a different equation of state for the gas
as presented by \citealt{roe05}). The second point to consider is that the
central regions of the disk in our simulation are too idealized, and so it is
hard to state how much of the inward flux created by the shock actually
reaches the centre of the galaxy. Also, the perfectly face-on geometry of the
interaction might have an influence of the shocked-gas galactic inflow. More
numerical experiments, with less idealized conditions, will be presented in
future contributions. Nevertheless, as long as the disk flares, oblique shocks
should appear for a face-on ICM wind interaction and the presence of a magnetic
field is a good mechanism to generate such a flare. Also, since a magnetized
disk is less compressible than a pure HD one, the shocked layer in the MHD
model will be more pressurized and will try to drain gas, either to the
outskirts and/or to the central regions of the disk.


\section{Conclusions} \label{conclusions}

We performed MHD and HD simulations of a disk galaxy subject to RPS
to analyze the impact of the MF in the dynamics of the gas during the
stripping event. Both models were set up in hydrostatic equilibrium
with the gravitational potential of an M33-like galaxy, without the
galactic bulge component of the potential.

We found that the galactic MF gives us a thicker gaseous disk than
the HD one, which change the dynamics of the model, that is, we have
gas farther from the galactic potential well (in the $z$ direction)
in the MHD, plus the surface density in $z$ is higher than in
{an HD disk with the same midplane density}.
When the ICM wind hits the disks, at the beginning of the
simulation, the MHD disk is hardly affected by the wind, since no
significant changes were observed in the initial shape of the disk,
only the compressed gas in the interaction interface. The HD disk is
easily perturbed and pushed off the galactic midplane by the wind.
Then the gravitational potential pulls back the material to disk,
generating an infall of gas to the disk until the ram pressure
exceeds the gravitational force and removes the gaseous disk.

The evolution of both models continues as the wind velocity
increases. Their ISM is removed of the disk, from the outside-in, and
reaches higher $z$ above the midplane.  When the models have evolved
for $t=500\Myr$, the swept gas in the MHD case is denser,
reaches a height of approximately $z \sim 20\kpc$, and the disk has
been truncated to $r\sim 10\kpc$. In the HD run, the swept gas is
{farther} away from the galactic midplane, $z\sim 40\kpc$, and has a
lower density than the MHD. The disk is also eroded to a smaller
radius of $r\sim 4-5\kpc$. These results {show} that the removal of
the gas disk is less efficient in the MHD model than in the HD case
{with the same midplane density}.

The main differences found so far between the models are: 

\begin{itemize}

  \item The HD disk is more easily {eroded} than the MHD one, because in
    the magnetized case we have a higher surface density $\Sigma$ and the gas is
    less compressible than in the HD model.  Since the surface density
    {strongly} affects the stripping rate, we developed an HD model with
    approximately the same $\Sigma$ as the MHD, {which shows} a similar
    stripping rate. This ``heavy'' HD disk has a very high midplane volumetric
    density that {makes it unrealistic}.

  \item The swept gas for the MHD model has a smooth appearance whilst for the
    HD models (both the regular and the heavy disks), the gas above the galaxy
    has a clumpier and filamentary-like morphology, that is, the MF mainly
    affects the shape and structure of the swept gas. 

\end{itemize}

Previous RPS simulations have obtained broader tails, that is the swept gas of
the disks, compared with observations of jellyfish galaxies (galaxies undergoing
RPS). It was expected that additional physical properties, such as MF, cooling,
star formation, etc. may help to solve this problem, presenting narrower tails
in the simulations. \cite{rus14} presented MHD simulations with radiative
cooling and self-gravity for a magnetized ICM only, and showed that the MF can
give narrower gas tails compared with HD models. Our runs show the opposite
behavior, the swept gas from the disk in the MHD model is broader than the HD,
but we do not have the same initial set-up as them. The differences in the tail
width could be also accounted for the radiative cooling. HD simulations performed
by \cite{ton10} including radiative cooling showed narrower tails in better
agreement with observations, compared to non-cooling models. On the other hand,
the swept gas from our MHD model shows a smooth structure, while the HD models
looks clumpier, si\-mi\-lar\-ly to tails observed in \cite{rus14}. {The
differences observed in the shape and morphology of the swept gas in our models
lie in the equation of state of the gas, that is an isothermal gas is more
compressible than an adiabatic (e.g. \citealt{roe05,roe07,roe08,ton09,ton10}) or
a magnetized gas, and when the wind hits the galaxy clump-like regions appear in
our HD simulations. When these regions are pushed and eroded by the wind, they
generate tails in the swept gas, where the flow is similar to the cigarette smoke,
giving the filamentary and clumpy shape to the swept gas in our HD simulation. A
more detailed analysis of the morphology and structure of the gas tails will be
presented in the near future (Ramos-Mart\1nez et al. in preparation).}
\cite{ton14} also performed RPS models with
galactic MF, with different configurations and intensities for the field, and
they found that the MFs do not make a significant difference in the stripping
rate of ISM, but the MF inhibits the mixing of the gas tail with the surrounding
ICM and unmixed gas survives at larger distances from the disk. In our results
we see a similar trend, since the swept gas in the MHD model also remains unmixed
for longer time, despite the fact that the $z$-height is smaller compared to our
HD run. The differences in the tail appearance and structure for their MHD and
HD models is not so evident or dramatic. Since the approach of our models is not
the same as \cite{rus14} and \cite{ton14} we cannot make an analytical
comparison with their works. We consider that, in order to understand if MF can
make a significant difference and its relevance in the interaction of the
ICM-ISM, further investigation will be needed.

Even when our HD simulation ran for $1\Gyr$, the model reached equilibrium at $t
\lesssim 500\Myr$: the truncated disk remained with the same radius although the
wind {was} still accelerating to a maximum velocity of $1000\kms$ before
the simulation ended. Therefore, we can assume that the MHD run has also reached
equilibrium with the ram pressure, or is near to it.  The remaining gaseous disk
could be removed by other mechanism, like interactions or fly-by's with other
galaxies (e.g. galaxy harassment), this should be taken into account because
these objects are not completely isolated, specially in clusters. Interactions
between galaxies can remove the gas or trigger star formation so the ISM is
consumed or exhausted.

It is well known that RPS works well removing the gas of the galaxies, but this
process fails in reproduce other S0s features, like higher bulge-to-disc ratios
than spirals, given that RPS has been proposed as a transformation mechanism of
spirals to S0s. For our magnetized case, {with inefficient RPS},
we found an interesting behavior in the gas: there are motions of gas from large
radii to the galactic {centre}. This phenomenon occurs only in the early
stages of the simulation, when the wind hits the disk, and it is produced by
oblique shocks at the interface of the interaction. The oblique shocks appear
because of our flared gas disk due to the MF presence and lead the gas to the
{centre} of the disk, which may help to maintain a reservoir of gas
available for star formation in the central region of the galaxy, which in
consequence could produce a thicker bulge that may lead to a higher
bulge-to-disk ratio. Studies have shown that the last star formation burst in
S0s galaxies took place in the bulge {(\citealt{pro11, sil06, sil12,
bed12, eve12, eve14}, but see \citealt{kat15})}.  Besides, if new stars are born
from the remaining gas in the {centre}, their strong winds could expel
the rest of the ISM from the galaxy.

Other observations of galaxies affected by RPS have shown unusual
nuclear activity, that is, the gas may be pushed to the
{centre} and the compression produced by the ICM enhances
star formation: the star formation is induced and enhanced by the
ram pressure \citep{cay90}. \cite{pog16} showed an atlas of
stripping candidates where most of their galaxies presented higher
star formation compared to non-stripped galaxies. From this
results, the oblique shocks can be seen as a mechanism that enhance
the formation of new stars in the remaining disk or even trigger
nuclear activity (e.g. an AGN). {Also \cite{pog17}, found a
very high incidence of AGN (Seyfert 2) among jellyfish galaxies
from MUSE data and the conclusion is that ram pressure triggers the
AGN activity.} Since the flux of gas derived from the oblique shocks
{in our MHD simulation} lasted about $\sim 150\Myr$
{from the time} the wind hit the disk, this could be
considered as comparable with the duty cycle of AGN's, which has
been estimated in $10-100\Myr$ \citep{hae93}, but given that our
{simulation does not properly model} the central regions of
the galaxy, {nor} we have a central black hole, we can only
{speculate} that the oblique shocks will transport the gas
for enough time to ignite an AGN. Other tests need to be
performed to better study the funneling of gas towards the
central regions of the galaxy, such as different wind profiles and
angles, and different disk surface densities and flare strengths.

Additionally, it has been reported that the star formation can continue in the
tail of the stripped gas, as it is shown in observations of HII regions in the
tail of a galaxy subject to RPS {\citep{keny99,bos06,cor07,yos08,hes10,sun10,
yag10,abr11,ken14,pog16}}. Due to the limitations of our
models, not enough resolution nor the appropriate equation of state to
solve the star formation, we cannot explore the possibility of new
stars born in the swept gas of our models or the centre of the disks
from the motions of gas originated from the oblique shocks.
More on this subject will be done in a future
work as well as an in-depth analysis of the swept gas for the MHD model (Ramos-Mart\1nez et al. in preparation).

\section*{Acknowledgments}

We thank B. Poggianti, M. Owers, and P. Appleton for useful comments and discussions.
We also thank an anonymous referee for comments that helped improve
this manuscript.
We acknowledge financial support from UNAM-DGAPA PAPIIT grant
IN100916, and CONACyT for support for MRM.





\begin{thebibliography}{}

\bibitem[Abadi et al.(1999)]{aba99}
  Abadi, M. G., Moore, B. \& Bower, R. G.
  1999, MNRAS, 308, 947

\bibitem[Abramson \& Kenney(2014)]{abr14}
  Abramson, A. \& Kenney, J. D. P.
  2014, AJ, 147, 63

\bibitem[Abramson et al.(2011)]{abr11}
  Abramson, A., Kenney, J. D. P., Crowl, H. H., Chung, A.,
van Gorkom, J. H., Vollmer, B. \& Schiminovich, D.
  2011, AJ, 141, 164

\bibitem[Abramson et al.(2016)]{abr16}
  Abramson, A., Kenney, J., Crowl, H., \& Tal, T.
  2016, AJ, 152, 32

\bibitem[Adebahr et al.(2013)]{ade13}
  Adebahr, B., Krause, M., Klein, U., We{\.z}gowiec, M.,
	Bomans, D. J. \& Dettmar, R. J.
  2013, A\&A, 555A, 23A

\bibitem[Aguerri et al.(2001)]{agu01}
  Aguerri, J., Balcells, M. \& Peletier, R.
  2001, A\&A, 428 

\bibitem[Aguerri et al.(2017)]{agu17}
  Aguerri, J., Agulli, I., Diaferio, A. \& Dalla Vecchia, C.
  2017, MNRAS, 468, 364

\bibitem[Allen \& Santill\'an(1991)]{potas91}
  Allen, C. \& Santill\'an, A.
  1991, Rev. Mex. Astron. Astrofis., 22, 255 

\bibitem[Beck(2005)]{bec05a}
  Beck, R. in {\textit{The magnetized plasma in galaxy evolution}}
  2005, ed. K. T. Chy{\.z}y, K. Otmianowska-Mazur, M. Soida \& R.-J. Dettmar, 193 

\bibitem[Beck et al.(2005)]{bec05b}
  Beck, R., Fletcher, A., Shukurov, A., Snodin, A., Sokoloff, D. D., Ehle, M., Moss, D. \& Shoutenkov, V.
  2005, A\&A, 444, 739B

\bibitem[Beck \& Wielebinski(2013)]{bec13}
  Beck, R. \& Wielebinski, R. in {\textit{Planets, Stars and Stellar Systems}}
  2013, Vol. 5, ed. G. Gilmore, Springer 

\bibitem[Bedregal(2012)]{bed12}
  Bedregal, A. G.
  2012, A\&AT, 27, 177

\bibitem[Bekki et al.(2002)]{bek02}
  Bekki, K., Couch, W. J. \& Shioya, Y.
  2002, ApJ, 577, 651

\bibitem[Bekki \& Couch(2003)]{bek03}
  Bekki, K. \& Couch, W. J.
  2003, ApJ, 596, L13 

\bibitem[Bekki(2014)]{bek14} Bekki, K.
  2014, MNRAS, 438, 444

\bibitem[Biviano \& Katgert(2004)]{biv04}
  Biviano, A. \& Katgert, P.
  2004, A\&A, 424, 779

\bibitem[Borlaff et al.(2014)]{bor14}
  Borlaff, A., Eliche-Moral, M. C., Rodr\1guez-P\'erez, C., Querejeta, M., Tapia, T.,
P\'erez-Gonz\'alez, P. G., Zamorano, J., Gallego, J. \& Beckman, J.
  2014, A\&A, 570, 103 

\bibitem[Boselli \& Gavazzi(2006)]{bos06}
  Boselli, A. \& Gavazzi, G.
  2006, PASP, 118, 517 

\bibitem[Boselli et al.(2014)]{bos14}
  Boselli, A., Cortese, L., Boquien, M., Boissier, S., Catinella, B., Gavazzi, G., Lagos, C. \& Saintonge, A.
  2014, A\&A, 564, A67

\bibitem[Butcher \& Oemler(1978)]{byo78}
  Butcher, H. \& Oemler, Jr., A.
  1978, ApJ, 219, 18 

\bibitem[Cayatte et al.(1990)]{cay90}
  Cayatte, V., van Gorkom, J. H., Balkowski, C. \& Kotanyi, C.
  1990, AJ, 100, 604

\bibitem[Chung et al.(2009)]{chu09}
  Chung, E. J., Rhee, M.-H., Kim, H., Yun, M. S., Heyer, M. \& Young, J. S.
  2009, ApJS, 184, 199

\bibitem[Chy{\.z}y \& Beck(2004)]{chy04}
  Chy{\.z}y, K. T. \& Beck, R.
  2004, A\&A, 417, 541C

\bibitem[Corbelli(2003)]{cor03}
  Corbelli, E.
  2003, MNRAS, 342,199 

\bibitem[Cortese et al.(2007)]{cor07}
  Cortese, L., Marcillac, D., Richard, J., Bravo-Alfaro, H., Kneib, J.-P.,
Rieke, G., Covone, G., Egami, E., Rigby, J., Czoske, O. \& Davies, J.
  2007, MNRAS, 376, 157

\bibitem[Dressler(1980)]{dre80}
  Dressler, A.
  1980, ApJ, 236, 351

\bibitem[Dressler(1986)]{dre86}
  Dressler, A.
  1986, ApJ, 301, 35 

\bibitem[Dressler et al.(1997)]{dre97}
  Dressler, A., Oemler, A., Smail, I., Barger, A., Butcher, H., Poggianti, B.
M. \& Sharples, R. M.
  1997, ApJ, 490, 577 

\bibitem[Elmegreen et al.(2002)]{elm02}
  Elmegreen, D. M., Elmegreen, B. G.,Frogel, J. A., Eskridge, P. B., Pogge, R. W.,
Gallagher, A. \& Iams, J.
  2002, AJ, 124, 777

\bibitem[Farouki \& Shapiro(1980)]{far80}
  Farouki, R. \& Shapiro, S. L.
  1980, ApJ, 241, 928

\bibitem[Fasano et al.(2000)]{fas00}
  Fasano, G., Poggianti, B. M., Couch, W. J., Bettoni, D, Kj{\ae}rgaard, P. \& Moles, M.
  2000, ApJ, 542, 673F

\bibitem[Fletcher(2010)]{fle10}
  Fletcher, A.
  2010, ASPC, 438, 197F

\bibitem[Fletcher et al.(2011)]{fle11}
  Fletcher, A., Beck, R., Shukurov, A., Berkhuijsen, E. M., Horellou, C.
  2011, MNRAS, 412, 2396

\bibitem[Fossati et al.(2016)]{fos16}
  Fossati, M., Fumagalli, M., Boselli, A., Gavazzi, G., Sun, M. \& Wilman, D. J.
  2016, MNRAS, 455, 2028

\bibitem[Frick et al.(2016)]{fri16}
  Frick, P, Stepanov, R, Beck, R, Sokoloff, D., Shukurov, A., Ehle, M. \& Lundgren, A.
  2016, A\&A, 585, A21

\bibitem[Fuchs \& von Linden(1998)]{fvl98}
  Fuchs, B. \& von Linden, S.
  1998, MNRAS, 294, 513

\bibitem[Fumagalli et al.(2014)]{fum14}
  Fumagalli, M., Fossati, M., Hau, G. K. T., Gavazzi, G., Bower, R., Sun, M. \& Boselli, A.
  2014, MNRAS, 445, 4335

\bibitem[Gallagher, Faber \& Balick(1975)]{gal75}
  Gallagher, J. S., Faber, S. M. \& Balick, B.
  1975, ApJ, 202, 7G

\bibitem[Gie{\ss}\"ubel(2012)]{gie12}
  Gie{\ss}\"ubel, R.
  2012, PhD Thesis, University of Cologne

\bibitem[Giraud(1986)]{gir86}
  Giraud, E.
  1986, A\&A, 167, 25

\bibitem[G\'omez \& Cox(2002)]{gom02}
  G\'omez, G. C. \& Cox, D.
  2002, ApJ, 580, 235 

\bibitem[Gunn \& Gott(1972)]{gun72}
  Gunn, J. E. \& Gott, J. R. I.
  1972, ApJ, 176, 1 

\bibitem[Haehnelt \& Rees(1993)]{hae93}
  Haehnelt, M. G. \& Rees, M. J.
  1993, MNRAS, 263, 168H

\bibitem[Hester et al.(2010)]{hes10}
  Hester, J. A., Seibert, M., Neill, J. D., Wyder, T K., Gil de Paz, A.,
Madore, B. F., Martin, D. C., Schiminovich, D. \& Rich, R. M.
  2010,ApJ, 716L, 14

\bibitem[Icke(1985)]{ick85}
  Icke, V.
  1985, A\&A, 144, 115 

\bibitem[J{\'a}chym et al.(2014)]{jac14}
  J{\'a}chym, P., Combes, F., Cortese, L., Sun, M. \& Kenney, J. D. P.
  2014, ApJ, 792, 11

\bibitem[J\'achym et al.(2009)]{jac09}
  J\'achym, P., K\"open, J., Palou\v{s}, J. \& Combes, F.
  2009, A\&A, 500, 693 

\bibitem[Johnston et al. (2012)]{eve12}
  Johnston E. J., Aragon-Salamanca A., Merrifield M. R. \& Bedregal A. G.
  2012, MNRAS, 422, 2590 

\bibitem[Johnston et al. (2014)]{eve14}
  Johnston E. J., Aragon-Salamanca A. \& Merrifield M. R.
  2014, MNRAS, 441, 333 

\bibitem[Kapferer et al.(2008)]{kap08}
  Kapferer, W., Kronberger, T., Ferrari, C., Riser, T. \& Schindler, S.
  2008, MNRAS, 389, 1405 

\bibitem[Katkov et al.(2015)]{kat15}
  Katkov, I. Y., Kniazev, A. Y. \& Sil'chenko, O. K.
  2015, AJ, 150, 24

\bibitem[Kenney \& Koopmann(1999)]{keny99}
  Kenny, J. D. P. \& Koopmann, R. A.
  1999, AJ, 117, 181 

\bibitem[Kenney et al.(2004)]{keny04}
  Kenney, J. D. P., van Gorkom, J. \& Vollmer, B.
  2004, AJ, 127, 3361 

\bibitem[Kenney et al.(2014)]{ken14}
  Kenney, J. D. P., Geha, M., J{\'a}chym, P., Crowl, H. H.,
Dague, W., Chung, A., van Gorkom, J. \& Vollmer, B.
  2014, ApJ, 780, 119

\bibitem[Klein et al.(1991)]{kle91}	
  Klein, U., Weiland, H. \& Brinks, E.
  1991, A\&A, 246, 323K

\bibitem[Koopmann \& Kenney(2004)]{koo04}
  Koopmann, R. A., \& Kenney, J. D. P.
  2004, ApJ, 613, 866

\bibitem[Kronberger et al.(2008)]{kron08}
  Kronberger, T., Kapferer, W., Ferrari, C., Unterguggenberger, S. \& Schindler, S.
  2008, A\&A, 481, 337 

\bibitem[Larson et al.(1980)]{lar80}
  Larson, R. B., Tinsley, B. M. \& Caldwell, C. N.
  1980, ApJ, 237, 692 

\bibitem[Moore et al.(1996)]{moo96}
  Moore, B., Katz, N., Lake, G., Dressler, A. \& Oemler, A.
  1996, Nat, 379, 613 

\bibitem[Niklas(1995)]{nik95}
  Niklas, S.
  1995, PhD Thesis, University of Bonn 

\bibitem[Oemler(1974)]{oem74}
  Oemler, A.
  1974, ApJ, 194, 10 

\bibitem[Otmianowska-Mazur \& Vollmer(2003)]{otm03}
  Otmianowska-Mazur, K. \& Vollmer, B.
  2003, A\&A, 402, 879 

\bibitem[Pfrommer \& Dursi(2010)]{pfr10}
  Pfrommer, C. \& Dursi, J.
  2010, Nature Physics, 6, 520 

\bibitem[Poggianti et al.(2016)]{pog16}
  Poggianti, B. M., Fasano, G., Omizzolo, A., Gullieuszik, M., Bettoni, D.,
Moretti, A., Paccagnella, A., Jaff\'e Y. L., Vulcani, B., Fritz, J,
Couch W. \& D'Onofrio, M.
  2016, AJ, 151, 78P

\bibitem[Poggianti et al.(2017)]{pog17}
  Poggianti, B. M., Jaff{\'e}, Y. L., Moretti, A., Gullieuszik, M., Radovich, M., and Tonnesen, S., 
Fritz, J., Bettoni, D., Vulcani, B., Fasano, G., Bellhouse, C., Hau, G. \& Omizzolo, A.
  2017, Nat, 548, 304P

\bibitem[Postman \& Geller(1984)]{pos84}
  Postman, M. \& Geller, M. J.
  1984, ApJ, 281, 95P

\bibitem[Prochaska Chamberlain et al.(2011)]{pro11}
  Prochaska Chamberlain, L. C., Courteau, S., McDonald, M \& Rose, J. A.
  2011, MNRAS, 412, 423P

\bibitem[Quilis et al.(2000)]{qui00}
  Quilis, V., Moore, B. \& Bower, R.
  2000, Science, 288, 1617 

\bibitem[Rengarajan et al.(1997)]{ren97}
  Rengarajan, T. N., Karnik, A. D. \& Iyengar, K. V. K.
  1997, MNRAS, 290, 1 

\bibitem[Roediger \& Br\"uggen(2006)]{ro06a}
  Roediger, E. \& Br\"uggen, M.
  2006, MNRAS, 369, 567 

\bibitem[Roediger \& Br\"uggen(2007)]{roe07}
  Roediger, E. \& Br\"uggen, M.
  2007, MNRAS, 380, 1399 

\bibitem[Roediger \& Br\"uggen(2008)]{roe08}
  Roediger, E. \& Br\"uggen, M.
  2008, MNRAS, 388, 465 

\bibitem[Roediger, Br\"uggen \& Hoeft(2006)]{ro06b}
  Roediger, E., Br\"uggen, M. \& Hoeft, M.
  2006, MNRAS, 371, 609 

 \bibitem[Roediger \& Hensler(2005)]{roe05}
  Roediger, E. \& Hensler, G.
  2005, A\&A, 433, 875 

\bibitem[Ruszkowski et al.(2014)]{rus14}
  Ruszkowski, M., Br\"uggen, M., Lee, D. \& Shin, M. S.
  2014, ApJ, 784, 75 

\bibitem[Schulz \& Struck(2001)]{sch01}
  Schulz, S. \& Struck, C.
  2001, MNRAS, 328,185 

\bibitem[Scodeggio \& Gavazzi(1993)]{sco93}
  Scodeggio, M. \& Gavazzi, G.
  1993, ApJ, 409, 110 

\bibitem[Sellwood \& Carlberg(1984)]{sel84}
  Sellwood, J. A. \& Carlberg, R. G.
  1984, ApJ, 282, 61

\bibitem[Sil'chenko(2006)]{sil06}
  Sil'Chenko, O.
  2006, ApJ, 641, 229

\bibitem[Sil'chenko et al.(2012)]{sil12}
  Sil'chenko, O.,  Proshina, I. S., Shulga, A. P. \& Koposov, S. E.
  2012, MNRAS, 427, 790

\bibitem[Simien \& de Vaucouleurs(1986)]{sim86}
  Simien, F. \& de Vaucouleurs, G.
  1986, ApJ, 302, 564S

\bibitem[Sivanandam et al.(2010)]{siv10}
  Sivanandam, S., Rieke, M. J. \& Rieke, G. H.
  2010, ApJ, 717, 147

\bibitem[Soida et al.(2006)]{soi06}
  Soida, M., Otmianowska-Mazur, K., Chy{\.z}y, K. \& Vollmer, B.
  2006, A\&A, 458, 727 

\bibitem[Stein et al.(2017)]{ste17}
  Stein, Y., Bomans, D. J., Ferguson, A. M. N. \& Dettmar, R.-J.
  2017, A\&A, 605, A5

\bibitem[Steinhauser et al.(2012)]{ste12}
  Steinhauser, D., Haider, M., Kapferer, W. \& Schindler, S.
  2012, A\&A, 544, A54 

\bibitem[Sun et al.(2010)]{sun10}
  Sun, M., Donahue, M., Roediger, E., Nulsen, P. E. J., Voit, G. M., Sarazin,
C., Forman, W. \& Jones, C.
  2010, ApJ, 708, 946 

\bibitem[Sun et al.(2007)]{sun07}
  Sun, M., Donahue, M. \& Voit, G. M.
  2007, ApJ, 671, 190

\bibitem[Sun et al.(2006)]{sun06}
  Sun, M., Jones, C. \& Forman, W.
  2006, ApJ, 637, L81 

\bibitem[Tabatabaei et al.(2008)]{tab08}
  Tabatabaei, F. S., Krause, M., Fletcher, A. \& Beck, R.
  2008, A\&A, 490, 1005

\bibitem[Tapia et al.(2014)]{tap14}
  Tapia, T. et al.
  2014, A\&A, 565, 31 

\bibitem[Teyssier(2002)]{tey02}
  Teyssier, R.
  2002, A\&A, 385, 337 

\bibitem[Tonnesen \& Bryan(2009)]{ton09}
  Tonnesen, S. \& Bryan, G. L.
  2009, ApJ, 694, 789 

\bibitem[Tonnesen \& Bryan(2010)]{ton10}
  Tonnesen, S. \& Bryan, G. L.
  2010, ApJ, 709, 1203 

\bibitem[Tonnesen \& Bryan(2012)]{ton12}
  Tonnesen, S. \& Bryan, G. L.
  2012, MNRAS, 422, 1609

\bibitem[Tonnesen \& Stone(2014)]{ton14}
  Tonnesen, S. \& Stone, J.
  2014, ApJ, 795, 148

\bibitem[Toomre \& Toomre(1972)]{too72}
  Toomre, A. \& Toomre, J.
  1972, ApJ, 178, 623 

\bibitem[Vollmer et al.(2001)]{vol01}
  Vollmer, B., Cayatte, V., Balkowski, C. \& Duschl, W. J.
  2001, ApJ, 561, 708 

\bibitem[Vollmer et al.(2004)]{vol04}
  Vollmer, B., Beck, R., Kenney, J. D. P. \& van Gorkom, J. H.
  2004, AJ, 127, 3375

\bibitem[Vollmer et al.(2008)]{vol08}
  Vollmer, B., Braine, J., Pappalardo, C. \& Hily-Blant, P.
  2008, \aap, 491, 455

\bibitem[Vollmer et al.(2000)]{vol00}
  Vollmer, B., Marcelin, M., Amram, P., Balkowski, C., Cayatte, V. \& Garrido, O.
  2000, A\&A, 364, 532

\bibitem[Vollmer et al.(2007)]{vol07}
  Vollmer, B., Soida, M., Beck, R., et al.
  2007, A\&A, 464, L37

\bibitem[Vollmer et al.(2006)]{vol06}
  Vollmer, B., Soida, M., Otmianowska-Mazur, K., Kenney, J. D., van Gorkom, J.
H. \& Beck, R.
  2006, A\&A, 453, 883 

\bibitem[Yagi et al.(2010)]{yag10}
  Yagi, M., Yoshida, M., Komiyama, Y.,  Kashikawa, N., Furusawa, H., Okamura, S.,
Graham, A. W., Miller, N. A., Carter, D., Mobasher, B. \& Jogee, S.
  2010, AJ, 140, 1814

\bibitem[Yoshida et al.(2008)]{yos08}
  Yoshida, M., Yagi, M., Komiyama, Y., Furusawa, H., Kashikawa,
N., Koyama, Y., Yamanoi, H., Hattori, T. \& Okamura, S.
  2008, ApJ, 688, 918-930

\bibitem[Zhang et al.(2013)]{zah13}
  Zhang, B., Sun, M., Ji, L., Sarazin, C., Lin, X. B., Nulsen, P. E. J., Roediger, E.,
Donahue, M., Forman, W., Jones, C., Voit, G. M. \& Kong, X.
  2013, ApJ, 777, 122 

\end{thebibliography}



\bsp	
\label{lastpage}
\end{document}